\documentclass[sigconf]{acmart}

\setcopyright{none}
\renewcommand\footnotetextcopyrightpermission[1]{}

\acmDOI{}
\acmISBN{}
\acmConference[][]{}
\acmBooktitle{}
\acmYear{}
\copyrightyear{}

\makeatletter
\def\@shortauthors{Preprint. Under review.}

\makeatother
\usepackage{multirow}
\usepackage{dblfloatfix}
\begin{document}
\fancyhf{}\fancyhead[LO,LE]{Preprint. Under review.}\fancyfoot[C]{\thepage}
\title[FOCAL: Preprint Under Review]{FOCAL: Filtered On-device Continuous Activity Logging for Efficient Personal Desktop Summarization}

\author{Haoran Yin}
\affiliation{%
  \institution{The Hong Kong Polytechnic University}
  \country{Hong Kong, China}
}
\email{hao-ran.yin@connect.polyu.hk}

\author{Zhiyuan Wen}
\affiliation{%
  \institution{The Hong Kong Polytechnic University}
  \country{Hong Kong, China}
}
\email{zyuanwen@polyu.edu.hk}

\author{Jiannong Cao}
\affiliation{%
  \institution{The Hong Kong Polytechnic University}
  \country{Hong Kong, China}
}
\email{jiannong.cao@polyu.edu.hk}

\author{Ruosong Yang}
\affiliation{%
  \institution{The Hong Kong Polytechnic University}
  \country{Hong Kong, China}
}
\email{rsong.yang@polyu.edu.hk}

\author{Bo Yuan}
\affiliation{%
  \institution{China Mobile Communications Company Limited Research Institute}
  \city{Beijing}
  \country{China}
}
\email{yuanbo@cmjt.chinamobile.com}



\begin{abstract}
Desktop interaction streams provide a continuous, privacy-sensitive record of interleaved user tasks. Transforming these streams into task-organized personal logs on-device faces two main challenges: exhaustive Vision-Language Model~(VLM) processing strains local resources, and global stream processing causes cross-task context pollution. We present \textbf{FOCAL} (\textbf{F}iltered \textbf{O}n-device \textbf{C}ontinuous \textbf{A}ctivity \textbf{L}ogging), a privacy-first multi-agent system utilizing a unified \emph{filter--plan--log} architecture. It cascades a lightweight Filter Agent for noise suppression, a text-only Brain Agent for task attribution, a Record Agent for selective visual reasoning, and a task-isolated Memory Agent for context-coherent summarization. Experiments on DesktopBench (comprising 2,572 screenshots across 420 complex sessions) show FOCAL reduces total token consumption by 60.4\% and VLM call count by 72.3\% versus a baseline, while boosting Key Information Recall~(KIR) from 0.38 to 0.61. Crucially, under $A{\to}B{\to}A$ task interruptions, FOCAL maintains Task~Acc~0.81 and KIR~0.80, whereas the baseline collapses to Task~Acc~0.03. FOCAL pioneers the efficient, on-device summarization of instruction-free desktop streams into multi-perspective personal logs.
\end{abstract}



%  \includegraphics[width=\textwidth]{Focal_new.png}
%  \caption{Overview of FOCAL.}
%  \Description{From desktop activity to personal structured log}


\maketitle

\begin{figure*}[t]
  \centering
  \includegraphics[width=0.9\textwidth]{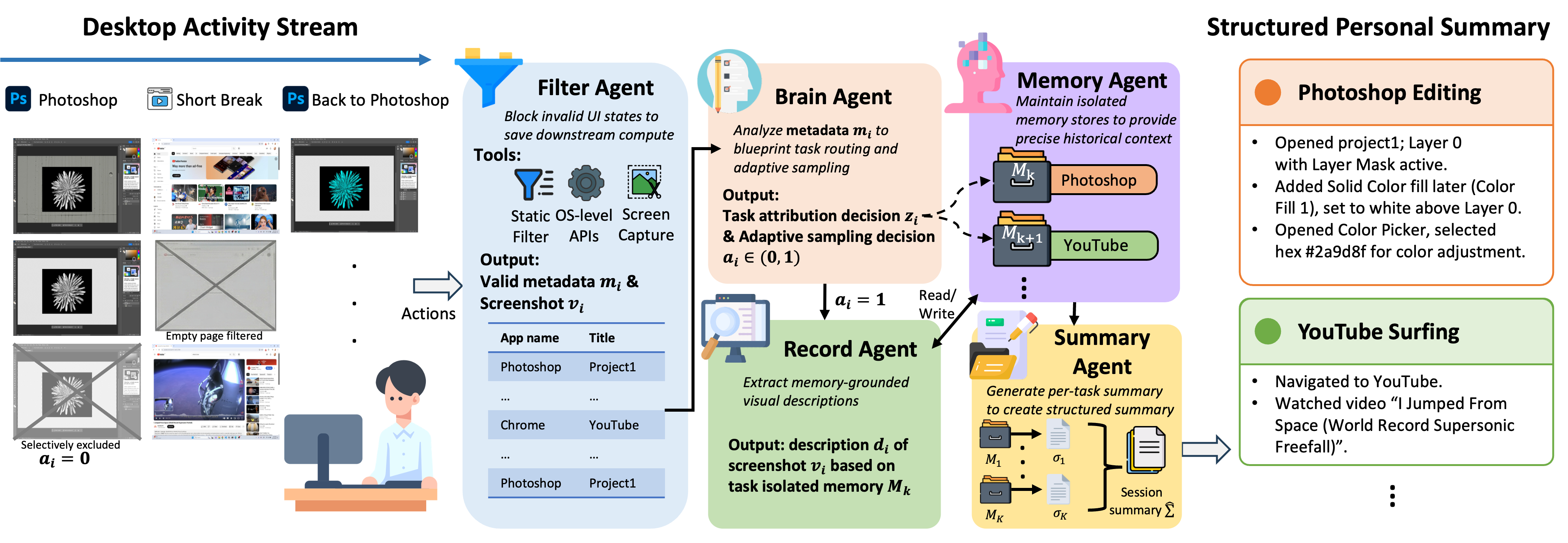}
  \caption{Overview of FOCAL.}
  \Description{FOCAL framework. From desktop activity to personal structured summary}
  \label{fig:focal}
\end{figure*}

\section{Introduction}
The information underlying personal computing is inherently multi-perspective: a single workday interleaves drafting documents, browsing references, debugging code, and communicating with collaborators into a continuous multimodal interaction stream \citep{gurrin2014lifelogging, sellen2010beyond, li2010stage}.
Turning this stream into compact, task-organized desktop logs could support productivity self-assessment and knowledge retrieval.
Desktop logging is also highly privacy-sensitive: screenshots and GUI traces may expose emails, chats, draft documents, credentials, and other sensitive content, making \textit{on-device understanding} a practical requirement.
Recent advances in large language models (LLMs) and vision-language
models (VLMs) have made semantic activity logging feasible for
wearable sensors \citep{he2025egolog} and mobile contexts
\citep{xu2025autolife}. However, they have not yet been extended to
privacy-first, on-device desktop interaction streams.

At the same time, existing progress in long-horizon multimodal reasoning, semantic distillation, and memory-based retrieval has substantially improved video understanding \citep{he2024ma, song2024moviechat, qian2024streaming, li2024llama, zou2024language, qiu2024mmsum, kim2024you, khanna2024goat}.
However, these approaches largely assume \textit{pre-collected offline inputs}, a \textit{given query or task label}, or deployment settings without strict on-device constraints.
A desktop workday, by contrast, is a continuous, instruction-free desktop interaction stream spanning multiple applications and interleaved tasks, where no task boundary is pre-annotated and no query is provided.
The missing capability is an instruction-free mechanism that can decide \textit{whether} a moment warrants visual reasoning and \textit{which} task context that observation should enter.

Bridging this gap exposes two intertwined challenges.
The first is \textbf{on-device inference intrusiveness}: every visual reasoning call competes with the user's foreground applications for local GPU, CPU, and memory resources, so exhaustive VLM processing introduces latency and resource contention.
The second is \textbf{cross-task context pollution}: users frequently switch among concurrent tasks, yet existing methods process the stream globally, causing shared memory to mix semantically similar but task-irrelevant descriptions.
Together, these challenges expose a critical gap: \textbf{existing methods lack an instruction-free pre-inference mechanism that both selects semantically valuable moments and routes evidence into task-specific context, forcing a trade-off between task-faithful reasoning and unobtrusive on-device operation.}
Addressing this gap is fundamentally a \textit{multi-perspective multimedia summarization} problem: each task provides a distinct semantic lens on the same continuous activity stream, and context-adaptive information access must be achieved without any prior query or user instruction.

To address this gap, we propose \textbf{FOCAL} (Filtered On-device Continuous Activity Logging), a privacy-first, on-device multi-agent logging system for desktop activity summarization that processes a continuous desktop interaction stream into multi-perspective, task-organized desktop logs through a unified ``filter-plan-log'' architecture.
The key idea is to move task-aware control \textit{before} expensive visual reasoning.
A lightweight \textit{Filter Agent} suppresses low-information events before any model call; a text-only \textit{Brain Agent} then uses GUI metadata to perform task attribution and adaptive planning; a VLM-based \textit{Record Agent} writes intent-centric log entries within task scope; a \textit{Memory Agent} enforces task-isolated context management; and a \textit{Summary Agent} composes task-level memories into coherent multi-perspective summaries.
By front-loading selection and routing, FOCAL reduces unnecessary local visual processing while preserving task-faithful context under continuous task switching.

To validate FOCAL, we conduct experiments on DesktopBench, reconstructed from VideoGUI, with both multi-task sessions and interruption scenarios.
We compare against a Naive LLM Agent (full-coverage processing without task-aware control) and FOCAL-GM (FOCAL with global shared memory), and evaluate both computational efficiency (VCC, TCS) and summary quality (BS-F1, Task Acc, KIR, G-Eval).
Results show that FOCAL effectively addresses the identified gap by jointly improving non-intrusive efficiency and task-faithful summarization in instruction-free desktop activity summarization.
In particular, FOCAL reduces visual token volume by 60.4\% relative to naive VLM inference while also improving summary quality.
We further perform ablation studies to verify the effectiveness of key local modules, especially adaptive sampling and task-isolated memory.

Our contributions are three-fold:
\begin{itemize}
\item \textbf{Privacy-First On-Device Logging.}
We propose FOCAL, a privacy-first, fully on-device multi-agent logging system that operationalizes filtered on-device continuous activity logging for instruction-free desktop activity summarization through a unified filter--plan--log architecture.
%We propose FOCAL, a ``filter-plan-log'' multi-agent architecture in which a lightweight Filter Agent and a text-only Brain Agent jointly perform task attribution and adaptive sampling \textit{before} any VLM call, enabling privacy-first and non-intrusive desktop logging while reducing visual token volume by 60.4\%.

\item \textbf{Dynamic Task-Switching Summarization.}
We introduce a task-isolated memory mechanism that routes
observations into per-task context stores. This structural isolation
prevents cross-task context pollution and enables coherent,
multi-perspective summaries under frequent task switching.
%We introduce a per-task memory architecture that routes each observation to a task-scoped store, structurally preventing cross-task context pollution and enabling coherent, multi-perspective personal logs even across long, heterogeneous work sessions with frequent task switching.

\item \textbf{Comprehensive Dataset and Evaluation.}
We construct DesktopBench, a session-level lifelogging benchmark curated from 2,572 raw GUI screenshots, comprising 320 compositional multi-task and 100 interruption sessions. Paired with a reproducible metric suite spanning efficiency (VCC, TCS) and summary quality (KIR, BS-F1, G-Eval), it establishes an annotation-efficient evaluation protocol for instruction-free continuous desktop logging.
\end{itemize}
\vspace{-6pt}
\section{Related Work}

\subsection{Personal Activity Logging}

Personal experience capture has long been studied in lifelogging \citep{gurrin2014lifelogging, sellen2010beyond}. Early systems such as MyLifeBits \citep{gemmell2002mylifebits} and SenseCam \citep{hodges2006sensecam} focused on continuous recording and retrieval, while newer work moves toward semantically readable personal narratives. This shift from raw capture to structured, language-mediated personal memory is especially important because logs become useful only when users can later search, skim, and reinterpret them at the level of daily activities and goals. AutoLife \citep{xu2025autolife}, for example, shows that LLMs can transform passive smartphone traces into structured daily journals without user input. At the same time, visual lifelogging remains inherently privacy-sensitive \citep{ferdous2017visualprivacy}, and recent MLLM studies further highlight the risk of sending personal visual data to cloud services \citep{hui2026mllmprivacy}. These concerns are even sharper for desktop logging, where screenshots may directly expose private documents, chats, and credentials.

However, existing personal logging systems rely mainly on physical-world signals and lack the application-layer semantics needed for desktop activities. They are therefore poorly suited to task-level understanding of interleaved GUI workflows spanning multiple windows, applications, and goals.

\subsection{Task-Oriented GUI Agents}

Recent GUI systems show that multimodal models can move beyond raw action detection toward semantic interface understanding. GUI-Narrator \citep{wu2024gui} generates natural-language descriptions of GUI actions, while task-oriented agents such as AutoDroid \citep{wen2024autodroid}, AppAgent v2 \citep{li2024appagent}, and UI-Hawk \citep{zhang2025ui} improve instruction-driven execution through structured histories, retrieval, or temporal modeling. Related long-context multimodal systems such as MovieChat \citep{song2024moviechat} and MA-LMM \citep{he2024ma} likewise highlight the importance of temporal memory.
Yet these systems are designed primarily for execution or query-conditioned reasoning rather than instruction-free retrospective logging. They usually assume explicit user goals, bounded trajectories, or global execution-centric memory. This assumption becomes problematic in real desktop workflows, where users frequently switch among applications and goals \citep{xie2024osworld, gottsacker2025xr}. In such settings, irrelevant context can accumulate and degrade reasoning quality \citep{shi2023large, liu2024lost}, motivating task-aware routing and isolated memory rather than a single shared context.

Overall, prior work leaves three gaps for our setting: desktop logging needs application-level task semantics, privacy-first on-device processing, and memory mechanisms that remain robust under frequent task switching. FOCAL targets these gaps as a multi-agent logging system for filtered on-device continuous activity logging and desktop summarization by combining pre-inference routing with task-isolated memory.

\section{Method}

\subsection{System Overview}

FOCAL is a multi-agent logging system for filtered on-device continuous activity logging: it converts a continuous desktop interaction stream into a task-organized session summary by moving task-aware control ahead of expensive visual reasoning. Here, an \emph{agent} denotes a module with restricted observations, outputs, and state access; the underlying LLM or VLM serves as the inference backbone of an agent rather than as an external tool.

As illustrated in Figure~\ref{fig:focal}, FOCAL comprises five agents with strictly separated responsibilities. \textit{Filter Agent} acquires valid actions, \textit{Brain Agent} performs a one-shot metadata-only planning pass, \textit{Record Agent} generates semantic descriptions only for sampled actions, \textit{Memory Agent} owns task-scoped persistent state, and \textit{Summary Agent} produces the final task-organized session summary. This separation is the central design principle of FOCAL: by deciding \emph{whether} to inspect a screenshot and \emph{which} task memory it should update before any VLM call, the system reduces unnecessary on-device visual computation while preserving task-faithful context under frequent task switching.

% ----------------------------------------------------------------
\subsection{Problem Formulation}
\label{sec:formulation}
% ----------------------------------------------------------------

We next formalize the input stream, the latent task structure, and the optimization target under the instruction-free on-device setting considered by FOCAL.

\paragraph{Given.}
We model a \textit{session} as an ordered desktop interaction sequence generated within one continuous working context, consisting of $N$ actions:
\begin{equation}
    \mathcal{S} = \{(m_i, v_i)\}_{i=1}^{N}
\end{equation}
where each action contains structured metadata
\begin{equation}
    m_i = (\text{app}_i,\ \text{title}_i)
\end{equation}
and a screenshot $v_i \in \mathbb{R}^{H \times W \times 3}$. A session contains $K$ latent \textit{tasks}; each action is associated with a task id $y_i \in [1, K]$. Define the action-index set of task $k$ as
\begin{equation}
    \mathcal{J}_k = \{i \in [1, N] \mid y_i = k\},
    \qquad
    T_k = \{(m_i, v_i)\}_{i \in \mathcal{J}_k}
\end{equation}
with partition constraints
\begin{equation}
    \bigcup_{k=1}^{K} \mathcal{J}_k = [1, N],
    \qquad
    \mathcal{J}_j \cap \mathcal{J}_k = \emptyset\ \ (j \neq k)
\end{equation}
The sets $\{\mathcal{J}_k\}$ partition the stream, but each $\mathcal{J}_k$ may be non-contiguous because users can suspend and resume tasks. The task count $K$ and assignments $\{y_i\}_{i=1}^{N}$ are latent and never provided as input.

\paragraph{Assume.}
We assume that actions are strictly time-ordered, task assignments and task count are unknown during processing, and all interaction data remains on-device throughout the pipeline.

\paragraph{Objective.}
Under these conditions, the goal is to generate a task-organized session-level summary $\hat{\Sigma}$ for session $\mathcal{S}$, defined as the ordered concatenation of per-task summaries:
\begin{equation}
    \hat{\Sigma} = \bigoplus_{k=1}^{\hat{K}} \hat{\sigma}_k
\end{equation}
where $\hat{K}$ is the number of tasks identified by the system, $\hat{\sigma}_k$ is the generated summary for task $k$, and $\oplus$ denotes ordered concatenation.
FOCAL jointly optimizes two objectives:
\begin{equation}
    \max_{\hat{\Sigma}}\ \text{Quality}(\hat{\Sigma},\ \Sigma^*)
    \quad \text{and} \quad
    \min\ \text{Cost}(\mathcal{S})
\end{equation}
Here, $\text{Quality}(\hat{\Sigma}, \Sigma^*)$ measures summary fidelity to the ground truth $\Sigma^*$, while $\text{Cost}(\mathcal{S})$ measures the computation consumed to process session $\mathcal{S}$. These objectives are coupled: higher fidelity often requires more VLM calls, whereas aggressive cost reduction risks missing important states.

\paragraph{Subject to.}
The key constraint is that FOCAL operates in an instruction-free regime: the input contains no user query, task labels, or task-boundary annotations. The Brain Agent plans only from metadata ($\text{app}$ and $\text{title}$), deciding for each action both task routing and whether visual sampling is necessary. Task identities are therefore induced by the system rather than supplied as supervision, distinguishing our setting from supervised task segmentation and query-conditioned video summarization.

\begin{table*}[!t]
\centering
\caption{Interfaces of FOCAL's five agents.}
\label{tab:agent_interfaces}
\footnotesize
\setlength{\tabcolsep}{2pt}
\begin{tabular}{p{0.08\textwidth} p{0.15\textwidth} p{0.17\textwidth} p{0.16\textwidth} p{0.16\textwidth} p{0.18\textwidth}}
\toprule
\textbf{Agent} & \textbf{Profile / Role} & \textbf{Observation} & \textbf{Output} & \textbf{State Access} & \textbf{Callable Interfaces} \\
\midrule
\textbf{Filter} & Front-end acquisition and validation module & Raw desktop environment state at the current timestamp & Valid action $(m_i, v_i)$ or discard decision & No persistent semantic state & Foreground metadata query; screenshot capture \\

\textbf{Brain} & Session-level planner for task routing and adaptive sampling & Full ordered metadata sequence $\{m_i\}_{i=1}^{N}$ & Session plan $\mathcal{P}$ and per-action decisions $(k_i, z_i, a_i)$ & Read-only metadata access; no screenshot or memory access & None beyond its own model inference \\

\textbf{Record} & Semantic recorder for sampled actions & Current screenshot $v_i$, metadata $m_i$, and task-local memory $\mathcal{M}_{k_i}^{(i)}$ & Memory-grounded description $\hat{d}_i$ & Read access to one task-local memory only & No arbitrary external tools \\

\textbf{Memory} & Stateful manager and sole owner of persistent cross-step task state & Task initialization, read, and append requests & Task memory snapshots and updated stores & Exclusive read-write access to $\{\mathcal{M}_k\}_{k=1}^{K}$ & \textsc{InitTask}, \textsc{ReadTask}, \textsc{AppendTask} \\

\textbf{Summary} & Session-end summarizer & Finalized isolated stores $\{\mathcal{M}_k\}_{k=1}^{\hat{K}}$ & Per-task summaries $\{\hat{\sigma}_k\}$ and session summary $\hat{\Sigma}$ & Read-only access to finalized task memories & None beyond its own model inference \\
\bottomrule
\end{tabular}
\end{table*}

% ----------------------------------------------------------------
\subsection{Agent Definitions and Interface Boundaries}
% ----------------------------------------------------------------

To make the decomposition precise, we distinguish \emph{agents} from \emph{tools}. In FOCAL, each agent is specified by what it may observe, what it may output, and what state it may access. Tools are explicit callable interfaces to external resources such as OS queries, screenshot capture, or memory APIs. Table~\ref{tab:agent_interfaces} summarizes these boundaries.

% ----------------------------------------------------------------
\subsection{FOCAL Processing Pipeline}
% ----------------------------------------------------------------

We next describe how these restricted agents are composed into a single end-to-end processing loop.

\paragraph{Action Acquisition and Validation.}
At each step, Filter Agent queries foreground metadata and captures the corresponding screenshot. Invalid observations are discarded immediately; otherwise, a valid action $(m_i, v_i)$ is emitted into the session stream. This stage ensures that downstream computation operates on a clean action sequence rather than raw desktop noise.

\paragraph{Session-level Planning.}
Once a session is formed, Brain Agent performs one metadata-only planning pass over the full ordered action list:
\begin{equation}
    \mathcal{P} = \text{BrainLLM}\!\left(\{m_i\}_{i=1}^{N}\right),
    \qquad
    m_i=(\text{app}_i,\ \text{title}_i)
\end{equation}
The resulting plan decouples control from perception. During sequential execution, each action is resolved into
\begin{equation}
    (k_i,\ z_i,\ a_i),
    \quad i=1,\ldots,N
\end{equation}
where $k_i$ is the resolved task id, $z_i$ indicates whether the action continues an existing task or initializes a new one, and $a_i \in \{0,1\}$ indicates whether visual sampling is required. If no valid task match exists, Memory Agent initializes a new task store. Thus, task routing and sampling are decided before any VLM call.

\paragraph{Conditional Recording and Memory Update.}
Visual reasoning is then performed only on sampled actions. When $a_i = 1$, Record Agent receives the screenshot $v_i$, metadata $m_i$, and the current task-local memory $\mathcal{M}_{k_i}^{(i)}$, and generates a memory-grounded semantic description:
\begin{equation}
    \hat{d}_i = \text{VLM}(v_i,\ m_i,\ \mathcal{M}_{k_i}^{(i)}),
    \quad \text{if } a_i = 1
\end{equation}
where $\mathcal{M}_{k_i}^{(i)} = \{\hat{d}_j\}_{j \in \mathcal{I}_{k_i}, j < i}$ is the set of previously generated descriptions for task $k_i$, and
\begin{equation}
    \mathcal{I}_k = \{j \in [1, N] \mid k_j = k,\ a_j = 1\}
\end{equation}
denotes the sampled-action index set of task $k$. The new description is not written directly by Record Agent; instead, Memory Agent appends it to the target store:
\begin{equation}
    \mathcal{M}_t^{(i+1)} =
    \begin{cases}
        \mathcal{M}_t^{(i)} \cup \{\hat{d}_i\}
        & \text{if } t = k_i,\ a_i = 1 \\
        \mathcal{M}_t^{(i)}
        & \text{otherwise}
    \end{cases},
    \quad \forall t \in [1, K_i]
\end{equation}
where $K_i = \left|\{k_j\mid 1\le j\le i\}\right|$ is the number of active tasks after action $i$. The key invariant is that each $\mathcal{M}_k$ contains only evidence routed to task $k$.

\paragraph{Per-task Summary Generation.}
After the final action has been processed, Summary Agent reads the finalized task stores and generates a structured summary for each task:
\begin{equation}
    \hat{\Sigma} = \bigoplus_{k=1}^{\hat{K}} \hat{\sigma}_k
\end{equation}
where $\hat{K}$ is the number of tasks identified by the system and $\oplus$ denotes ordered concatenation. Because summarization is deferred until the end of the session, each task summary is generated from the complete evidence accumulated for that task.

% ================================================================
\section{DesktopBench Dataset Construction}
\label{sec:dataset_construction}
% ================================================================
Instruction-free desktop activity summarization requires session-level benchmarks that preserve cross-application workflows and task continuity under switching, which standard single-task GUI benchmarks do not provide.
We reconstruct \textbf{VideoGUI}~\citep{lin2024videogui} into \textbf{DesktopBench}, a benchmark for multi-task desktop activity logging. Following Table~\ref{tab:task_family}, we regroup its source tasks into five application families spanning editing, generation, slides, and reference activities.
DesktopBench has two splits: \textbf{DesktopBench-Multitask} for interleaved cross-application workflows and \textbf{DesktopBench-Interruption} for controlled $A\!\rightarrow\!B\!\rightarrow\!A$ timelines that test whether a method can preserve and resume task state after interruption.

\begin{table}[t]
\centering
\caption{Task family taxonomy of DesktopBench.}
\label{tab:task_family}
\small
\resizebox{\columnwidth}{!}{%
\begin{tabular}{llll}
\toprule
\textbf{Family} & \textbf{Description} & \textbf{Prefix} & \textbf{Application} \\
\midrule
\multirow{4}{*}{Video Editing} & \multirow{4}{*}{Timeline, effects, export} & AE & Adobe After Effects \\
 & & PR & Adobe Premiere Pro \\
 & & DV & DaVinci Resolve \\
 & & CC & CapCut \\
\midrule
Image Editing & Asset processing & PS & Adobe Photoshop \\
\midrule
\multirow{4}{*}{Generation} & \multirow{4}{*}{AI multimedia creation} & RW & Runway \\
 & & SD & Stable Diffusion WebUI \\
 & & DALLE & DALL\textperiodcentered E \\
 & & AI & Other AI tools \\
\midrule
Slides & Presentation authoring & PPT & Microsoft PowerPoint \\
\midrule
\multirow{3}{*}{Reference} & \multirow{3}{*}{Tutorials, asset browsing} & WEB & Web browser \\
 & & YT & YouTube \\
 & & VLC & VLC Media Player \\
\bottomrule
\end{tabular}%
}
\end{table}

\begin{figure*}[t]
  \centering
  \includegraphics[width=0.85\textwidth]{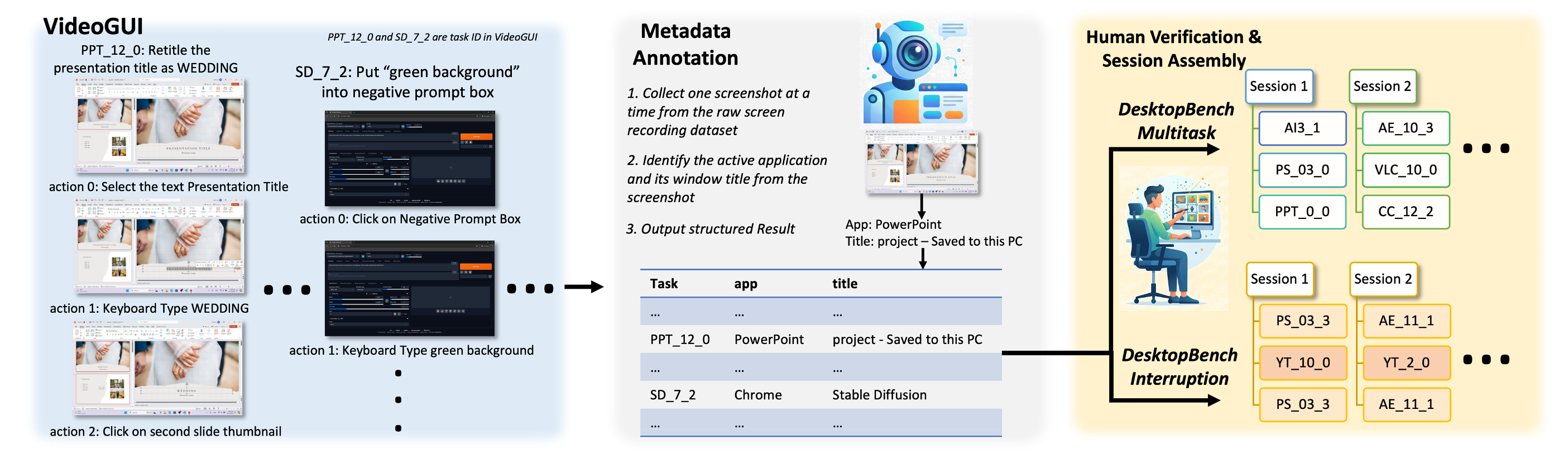}
  \caption{DesktopBench construction pipeline with Multi-task and Interruption splits.}
  \Description{Diagram showing the dataset construction pipeline from original VideoGUI subtasks to Multi-task and Interruption session splits.}
  \label{fig:dataset_construction}
\end{figure*}

\subsection{Session Construction}
\label{sec:session_construction}
Because VideoGUI was built for single-task automation, it lacks the session organization and metadata required for activity logging. We therefore augment each action with the foreground application name (\texttt{app}) and window title (\texttt{title}), while retaining the original task descriptions as semantic reference (Figure~\ref{fig:dataset_construction}).

\textit{DesktopBench-Multitask (320 sessions).}
Sessions are assembled through template-based composition grounded in realistic creative workflows. We use 20 patterns spanning video-centric workflows such as \texttt{video$\to$ref$\to$video} and design-centric workflows such as \texttt{generation$\to$image$\to$slide}. Each session contains at least two distinct application prefixes, and compatible subtasks may be reused to compose richer workflows.
\textit{DesktopBench-Interruption (100 sessions).}
To evaluate robustness to context switching, we construct an $A\!\rightarrow\!B\!\rightarrow\!A$ split in which a long-running creative task~$A$ is interrupted by a short YouTube browsing task~$B$ before resuming. This tests whether a model can preserve and recover the latent state of task~$A$ without cross-task drift.

\subsection{Ground-Truth Summary Construction}
\label{sec:groundtruth}
To obtain reference summaries independent of model-predicted segmentation, we construct ground truth from human-annotated task assignments $\{\mathcal{J}_k^*\}_{k=1}^{K^*}$ and the original task descriptions. GPT-5~\citep{openai2026gpt5} first produces frame-level visual descriptions, which are then aggregated within each ground-truth task into task-level reference summaries.
This anchors references to annotated task identities rather than model-predicted ones, decoupling summarization fidelity from segmentation accuracy and making wrong-task attribution or cross-task leakage easier to expose during evaluation.

\subsection{Dataset Statistics}
\label{sec:dataset_stats}
\begin{figure*}[t]
  \centering
  \includegraphics[width=0.96\textwidth]{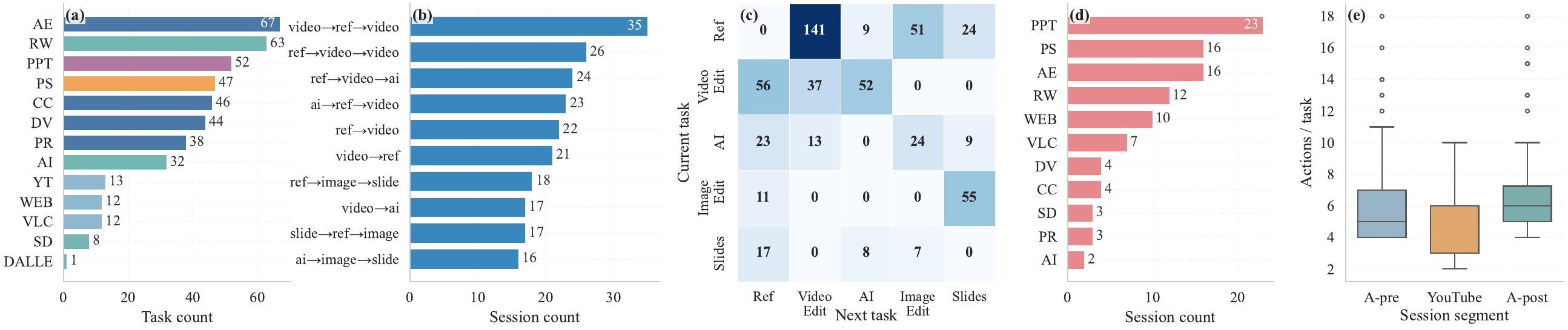}
  \caption{DesktopBench statistics. (a)~Task prefix counts (Multi-task). (b)~Session pattern distribution. (c)~Task-type transitions. (d)~Interrupted task distribution. (e)~Actions per segment (Interruption).}
  \Description{Five-panel figure showing dataset statistics: task prefix distributions, session composition patterns, task-type transitions for Multi-task, and interrupted task distributions and per-segment action counts for Interruption.}
  \label{fig:dataset_overview}
\end{figure*}

DesktopBench is reconstructed from 2,572 screenshots and uses the \textit{session} as the evaluation unit. It contains 420 sessions in total: 320 in DesktopBench-Multitask and 100 in DesktopBench-Interruption. Multitask sessions contain 2--4 tasks, while Interruption sessions contain exactly two tasks under the $A\!\rightarrow\!B\!\rightarrow\!A$ structure.
Table~\ref{tab:dataset} summarizes the key statistics. The average session length is 17.3 actions for DesktopBench-Multitask and 16.5 for DesktopBench-Interruption. Because compatible source subtasks may be reused when composing workflows, these counts measure instantiated actions with multiplicity rather than unique screenshots. Figure~\ref{fig:dataset_overview} visualizes the two splits.

\begin{table}[h]
\centering
\caption{Dataset statistics of DesktopBench.}
\label{tab:dataset}
\small
\begin{tabular}{p{0.50\linewidth}cc}
\toprule
\textbf{Statistic} & \textbf{Multitask} & \textbf{Interruption} \\
\midrule
Number of sessions & 320 & 100 \\
Average session length (actions) & 17.3 & 16.5 \\
Task count per session & 2--4 & 2 \\

\bottomrule
\end{tabular}
\end{table}

% ================================================================
\section{Experiment}
% ================================================================
\subsection{Baselines}
\label{sec:baselines}

We compare FOCAL against two baselines, each designed to ablate a core component of the proposed pipeline.

\paragraph{Naive LLM Agent.}
This baseline invokes the VLM on every action in the session to produce a context-free visual description, without any sampling filter or task memory.
All descriptions are concatenated chronologically and passed to an LLM that generates a single session-level summary with task boundary awareness.
By removing both adaptive sampling and task-isolated memory, this baseline provides a direct reference for the efficiency--quality trade-off.

\paragraph{FOCAL-GM (Global Memory).}
FOCAL-GM retains the Brain Agent and its adaptive sampling mechanism,
but replaces task-isolated memory with one global memory pool shared
across all tasks.
When the VLM generates a description, it retrieves context from this shared pool rather than from a task-specific store.
The rest of the pipeline---task attribution and sampling decisions---remains identical to FOCAL.
This baseline isolates the contribution of the task-isolation mechanism while holding the sampling strategy constant.

We do not compare against existing personal logging systems such as AutoLife~\cite{xu2025autolife} or EgoLog~\cite{he2025egolog}, as these systems operate on fundamentally different input modalities (smartphone sensor traces and wearable signals, respectively) and assume a single continuous activity rather than interleaved multi-task desktop workflows. Adapting them to the instruction-free, multi-task GUI stream setting considered here would require non-trivial re-engineering that goes beyond fair baseline comparison; our work instead targets a previously unaddressed problem regime. 
We likewise do not include rule-based or heuristic task-segmentation baselines (e.g., window-change triggers or application-prefix matching), as the heterogeneous and context-dependent nature of desktop metadata where the same application prefix can span semantically unrelated tasks and a single task can span multiple applications makes it impossible to define a unified segmentation rule that generalizes across the full diversity of sessions in
DesktopBench. 
The two ablation baselines above are therefore designed to isolate the specific contributions of adaptive sampling and task-isolated memory within a controlled pipeline, which is the primary evaluation objective of this work.

\subsection{Evaluation Metrics}
\label{sec:metrics}

We evaluate FOCAL along two complementary dimensions: computational efficiency and summary quality.

\subsubsection{Computational Efficiency}

\paragraph{VLM Call Count (VCC).}
VCC measures the average number of VLM invocations per session over $S$ sessions, directly quantifying the efficiency gain from adaptive sampling:
\begin{equation}
    \text{VCC} = \frac{1}{S} \sum_{s=1}^{S} \sum_{k=1}^{\hat{K}} |\mathcal{I}_k|
\end{equation}
where $\mathcal{I}_k = \{i \in [1, N] \mid k_i = k,\ a_i = 1\}$ denotes the set of frames assigned to predicted task~$k$ and selected for VLM processing, and $\hat{K}$ is the total number of predicted tasks at the end of session~$s$.

\paragraph{Token Consumption per Session (TCS).}
TCS captures the total token budget consumed by all model calls when processing one session:
\begin{equation}
    \text{TCS} = \tau_{\text{brain}}
    + \tau_{\text{record}}
    + \tau_{\text{sum}}
\end{equation}
where $\tau_{\text{brain}}$, $\tau_{\text{record}}$, and $\tau_{\text{sum}}$ denote the cumulative token consumption of the Brain Agent, Record Agent, and Summary Agent over the entire session, respectively.

\begin{table*}[h]
\centering
\caption{Token efficiency on the Multi-task and Interruption splits.}
\label{tab:efficiency_results}
\resizebox{\textwidth}{!}{%
\begin{tabular}{lccccc|ccccc}
\toprule
& \multicolumn{5}{c|}{Multi-task} & \multicolumn{5}{c}{Interruption} \\
Method & VCC & TCS $\downarrow$ & Brain Tok. $\downarrow$ & VLM Tok. $\downarrow$ & Sum. Tok. $\downarrow$ & VCC & TCS $\downarrow$ & Brain Tok. $\downarrow$ & VLM Tok. $\downarrow$ & Sum. Tok. $\downarrow$ \\
\midrule
Naive LLM Agent & 17.3 & 45{,}240 & - & 42{,}583 & 2{,}657 & 16.5 & 45{,}178 & - & 40{,}775 & 4{,}403 \\
FOCAL-GM & \textbf{4.3} & 21{,}260 & 3{,}605 & 15{,}670 & 1{,}986 & \textbf{4.3} & 21{,}557 & \textbf{4{,}340} & 15{,}666 & 1{,}551 \\
FOCAL & 4.8 & \textbf{17{,}913} & \textbf{3{,}744} & \textbf{12{,}873} & \textbf{1{,}296} & 4.4 & \textbf{17{,}408} & 4{,}410 & \textbf{11{,}592} & \textbf{1{,}406} \\
\bottomrule
\end{tabular}%
}
\end{table*}

\subsubsection{Summary Quality}

\paragraph{Key Information Recall (KIR).}
KIR assesses whether the key information from each ground-truth task is preserved in the semantically matched predicted summaries.
For each GT task $k \in [1, K^*]$, a judge model identifies the set of semantically matched predictions $\mathcal{A}(k) \subseteq \{\hat{\sigma}_1, \dots, \hat{\sigma}_{\hat{K}}\}$.
Every GT key point then falls into one of three categories: correctly recalled in a matched summary ($c$), attributed to a wrong task ($w$), or missing entirely ($m$).
KIR aggregates these counts as
\begin{equation}
    \mathrm{KIR} = \frac{c}{c + \lambda w + m}
\end{equation}
where $\lambda$ controls the penalty for wrong-task attribution.
We set $\lambda\!=\!2$, which guarantees $\mathrm{KIR} \in [0, 1]$ and penalizes cross-task leakage more heavily than outright omission.

\paragraph{Task Count Accuracy (Task Acc).}
Task Acc measures the fraction of sessions for which the predicted number of tasks exactly matches the ground truth:
\begin{equation}
\mathrm{Task\ Acc}
=
\frac{1}{S}
\sum_{i=1}^{S}
\mathbf{1}\!\left({\hat{K}}_i = K_i^*\right),
\end{equation}
where $\hat{K}_i$ and $K_i^*$ denote the predicted and ground-truth task counts for session~$i$, and $\mathbf{1}(\cdot)$ is the indicator function.

\paragraph{BERTScore F1 (BS-F1).}
BS-F1 measures the semantic similarity between a generated summary and the ground truth via contextual token embeddings from a pre-trained language model, providing robustness to surface-level paraphrase variation:
\begin{equation}
    \text{BS-F1} =
    \frac{2 \cdot \text{BS-P} \cdot \text{BS-R}}
         {\text{BS-P} + \text{BS-R}}
\end{equation}
where BS-P and BS-R are BERTScore precision and recall computed over token-level contextual embeddings between $\hat{\Sigma}$ and $\Sigma^*$.

\paragraph{G-Eval.}
Following~\citet{openai2026gpt5}, we employ GPT-5 as a judge to score each generated summary on five dimensions: Accuracy (factual consistency with the interaction trace), Coverage (completeness of key task information), Conciseness (absence of redundant content), Consistency (internal logical coherence), and Clarity (readability and fluency).
The overall G-Eval score is the mean across dimensions:
\begin{equation}
    \text{G-Eval} = \frac{1}{5}
    \sum_{d \in \mathcal{D}} \text{score}_d(\hat{\Sigma})
\end{equation}
where $\mathcal{D}$ contains five dimensions---Accuracy, Coverage,
Conciseness, Consistency, and Clarity---and each score lies on a 1--5
scale.
G-Eval complements BS-F1 by detecting logical errors and hallucinated content that embedding-based metrics cannot capture.

\begin{figure}[t]
  \centering
  \includegraphics[width=0.7\columnwidth]{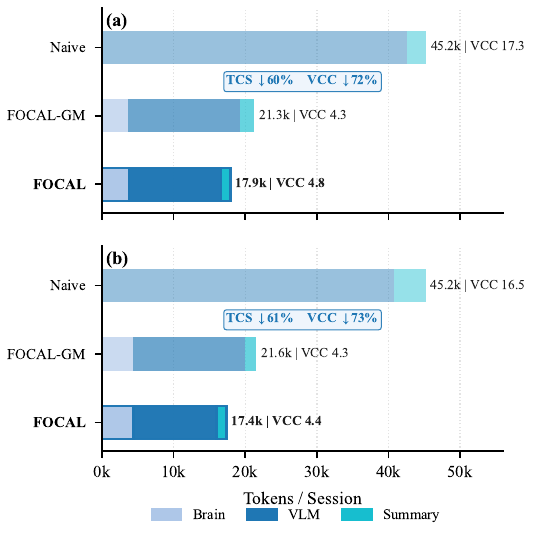}
  \caption{Token efficiency on the Multi-task and Interruption splits.}
  \Description{Single-column figure with two horizontal stacked-bar panels: panel (a) for Multi-task and panel (b) for Interruption. Each panel compares Naive, FOCAL-GM, and FOCAL. Bars are segmented into Brain, VLM, and Summary tokens, and text labels show total token count and VCC. Blue callouts summarize FOCAL's percentage reductions relative to the naive baseline.}
  \label{fig:efficiency}
\end{figure}

\section{Results}
All experiments are conducted on a MacBook with an Apple M4 chip and 16\,GB unified memory, reflecting the resource-constrained on-device setting that FOCAL targets. The language model used throughout is \texttt{qwen3:8b}, a compact 8-billion-parameter model chosen to validate that the proposed pipeline can achieve competitive performance without relying on large-scale cloud-hosted models.

\subsection{Token Efficiency}

%Table~\ref{tab:efficiency_results} presents the computational efficiency of FOCAL relative to both baselines on the DesktopBench-Multitask and DesktopBench-Interruption datasets. On the multi-task dataset, FOCAL reduces VCC from 17.3 to 4.8 calls per session ($-72.3\%$), TCS from 45{,}240 to 17{,}913 tokens per session ($-60.4\%$), and VLM tokens from 42{,}583 to 12{,}873 ($-69.8\%$). The summary token cost is likewise reduced from 2{,}657 to 1{,}296 ($-51.2\%$), since task-isolated memories yield more focused inputs for the Summary Agent. Compared with FOCAL-GM, which already benefits from Brain Agent--driven adaptive sampling, FOCAL further lowers TCS by 15.7\% and VLM tokens by 17.8\%. This additional gain is attributable to task-isolated memory: by supplying only within-task context to the Record Agent, the system avoids accumulating redundant cross-task descriptions that inflate prompt length. The dominant cost reduction originates from VLM tokens, which account for over 70\% of TCS in all configurations, confirming that metadata-guided sampling effectively suppresses unnecessary visual reasoning. On the interruption dataset, FOCAL maintains comparable efficiency (VCC 4.4, TCS 17{,}408), indicating that the pipeline's computational overhead remains stable even under frequent context switches characteristic of $A{\to}B{\to}A$ workflows. Figure~\ref{fig:efficiency} summarizes these end-to-end cost savings while preserving the token-component breakdown.

Table~\ref{tab:efficiency_results} presents the computational efficiency of FOCAL relative to both baselines. 
On the multi-task dataset, FOCAL reduces VCC by 72.3\% (17.3$\to$4.8 calls/session) and TCS by 60.4\% (45{,}240$\to$17{,}913 tokens/session) versus the naive baseline, while cutting VLM tokens by 69.8\%. 
Compared with FOCAL-GM, FOCAL further lowers TCS by 15.7\%, showing that task-isolated memory suppresses redundant cross-task context without increasing sampling cost.
On the interruption dataset, FOCAL maintains similar efficiency (VCC~4.4, TCS~17{,}408), indicating stable overhead under $A{\to}B{\to}A$ task switching; Figure~\ref{fig:efficiency} shows the token breakdown.

\begin{table}[h]
\centering
\footnotesize
\setlength{\tabcolsep}{3pt}
\caption{Ablation of the Brain Agent and memory strategy on the multi-task split.}
\label{tab:ablation_agent_memory}
\begin{tabular}{llccccc}
\toprule
Brain & Memory & TCS $\downarrow$ & BS-F1 $\uparrow$ & Task Acc $\uparrow$ & KIR $\uparrow$ & G-Eval $\uparrow$ \\
\midrule
\multirow{2}{*}{w/o} & No Memory    & 45{,}420 & \textbf{0.866} & 0.72 & 0.38 & 2.96 \\
                     & Global Memory & 47{,}625 & 0.869          & \textbf{0.79} & 0.66 & 3.47 \\
\midrule
\multirow{3}{*}{w/}  & No Memory                       & 18{,}759 & 0.854 & 0.68 & 0.36 & 2.87 \\
                     & Global Memory                   & 21{,}261 & 0.855 & 0.68 & 0.42 & 3.41 \\
                     & Task-Isolated  (FOCAL)    & \textbf{17{,}913} & 0.857 & 0.69 & \textbf{0.61} & \textbf{4.16} \\
\bottomrule
\end{tabular}
\end{table}

\subsection{Summary Quality}

Table~\ref{tab:quality_results} reports summary quality across four metrics. 
On the multi-task dataset, FOCAL achieves KIR~0.61 and G-Eval~4.16, improving over the naive baseline by +0.23 and +1.20 respectively, and over FOCAL-GM by +0.19 and +0.75---gains that arise entirely from task-isolated memory preventing cross-task context pollution. 
Notably, the naive baseline achieves the highest BS-F1 (0.866) despite the lowest KIR and G-Eval, exposing a fundamental limitation of embedding-based overlap metrics in multi-task settings: a summary that mixes content across tasks can still exhibit high token-level similarity to the concatenated ground truth while violating task-level semantic boundaries. 
KIR and G-Eval are therefore more faithful indicators of actionable desktop logs.

\begin{figure}[t]
  \centering
  \includegraphics[width=0.9\columnwidth]{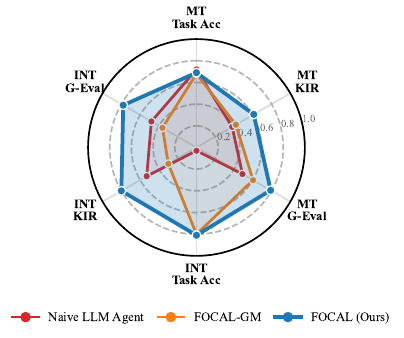}
  \caption{Summary quality on the Multi-task and Interruption splits.}
  \Description{Radar chart with two overlaid polygons representing the Multi-task (MT) and Interruption (INT) splits. Each radar has three axes: Task Accuracy, KIR, and G-Eval (normalized to [0,1]). Three methods are compared---Naive LLM Agent, FOCAL-GM, and FOCAL---shown as concentric polygons of increasing area. FOCAL consistently occupies the largest area on both splits, with blue delta labels annotating its improvement over the naive baseline on each axis.}

  \label{fig:quality}
\end{figure}

On the interruption dataset, the naive baseline collapses to Task Acc~0.03 under $A{\to}B{\to}A$ switching, whereas FOCAL maintains Task Acc~0.81, KIR~0.80, and G-Eval~4.12, confirming robust task attribution under stream fragmentation.
Figure~\ref{fig:quality} visualizes these comparisons across both datasets.

\subsection{Ablation Study: Agent Component}
\label{sec:ablation_study}

Table~\ref{tab:ablation_agent_memory} reports five configurations that progressively add the Brain Agent and memory isolation on the multi-task dataset, revealing two complementary contributions.

The Brain Agent drives the main efficiency gain: removing it (rows~1 vs.~3) raises TCS from 18{,}759 to 45{,}420 ($+58.7\%$), confirming that one-shot metadata planning suppresses redundant VLM invocations. 
However, sampling alone does not improve quality: without memory, KIR and G-Eval slightly \emph{decrease} (0.38$\to$0.36, 2.96$\to$2.87) because discarded frames remove useful evidence.

\begin{figure}[t]
  \centering
  \includegraphics[width=0.9\columnwidth]{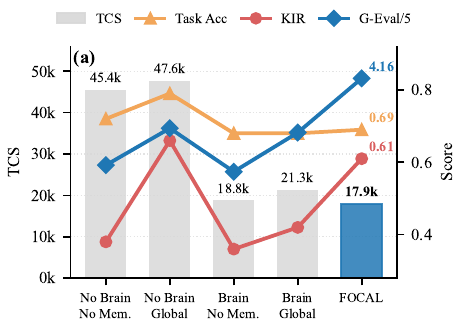}
  \caption{Ablation of the Brain Agent and memory strategy on the multi-task split.}

 \Description{Combined bar-and-marker ablation chart with a shared x-axis of five configurations: No Brain/No Memory, No Brain/Global Memory, Brain/No Memory, Brain/Global Memory, and FOCAL (Brain/Task-Isolated Memory). Gray bars represent token consumption per session (TCS) read on the left axis, ranging from 17.9k to 47.6k. Three markers overlaid on the bars track Task Accuracy, KIR, and G-Eval (normalized, right axis) across configurations. FOCAL achieves the lowest TCS at 17.9k while reaching the highest scores on all three quality metrics.}

  \label{fig:ablation}
\end{figure}

Memory isolation provides the key quality recovery. Progressing from no memory to global to task-isolated (rows~3$\to$4$\to$5), KIR rises from 0.36 to 0.42 to 0.61 and G-Eval from 2.87 to 3.41 to 4.16, while TCS remains stable. 
The global$\to$isolated transition is especially informative: the +0.19 KIR and +0.75 G-Eval gains come entirely from eliminating cross-task contamination, with no change in sampling strategy. 
Row~2's highest BS-F1 (0.869) again shows that embedding-based overlap is insensitive to task-boundary violations, making KIR and G-Eval the more reliable quality indicators.
Figure~\ref{fig:ablation} visualizes this monotonic trajectory.

\begin{figure*}[!t]
  \centering
  \includegraphics[width=0.9\textwidth]{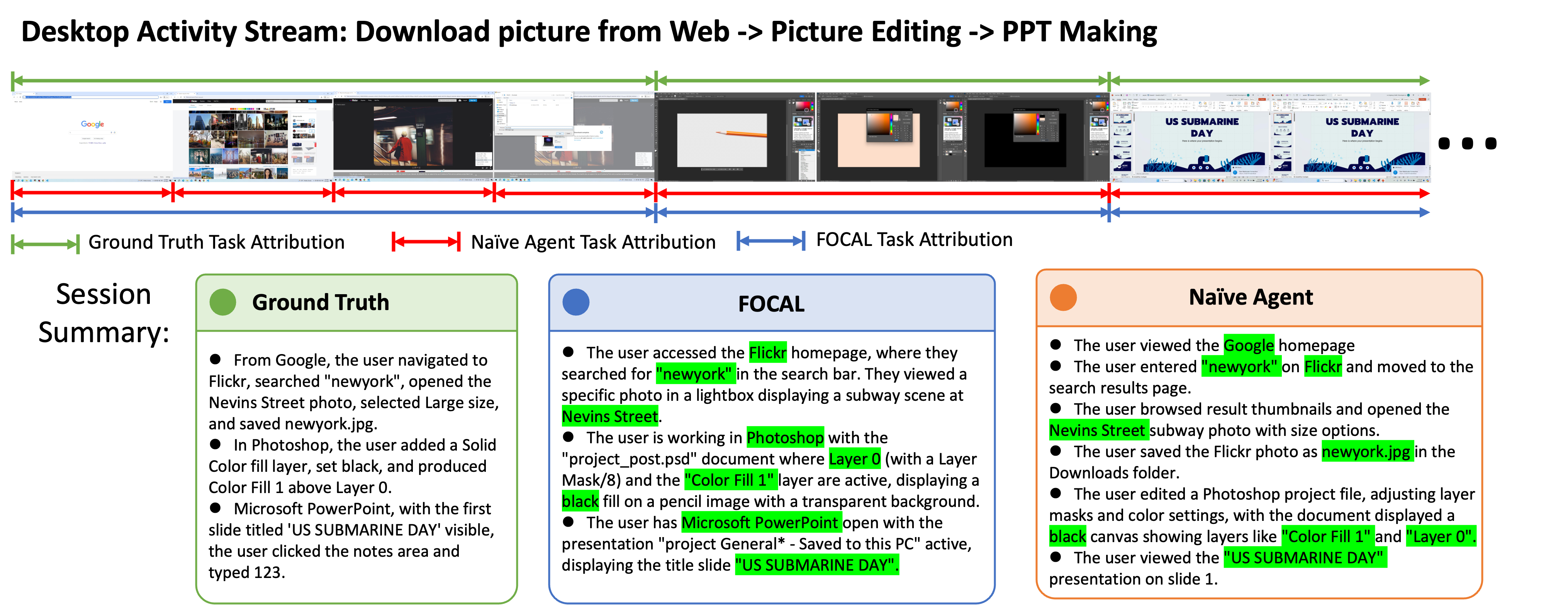}
  \caption{Case study of Session~258. FOCAL recovers three task spans, while the naive baseline over-segments the workflow.}
  \Description{Case study figure showing a raw desktop activity stream with screenshots and comparing ground-truth task spans, FOCAL task spans, and naive agent segments for Session 258.}
  \label{fig:case_study_stream}
\end{figure*}

\begin{table*}[h]
\centering
\small
\caption{Summary quality on the Multi-task and Interruption splits.}
\label{tab:quality_results}
%\resizebox{\textwidth}{!}{%
\begin{tabular}{lcccc|cccc}
\toprule
& \multicolumn{4}{c|}{Multi-task} & \multicolumn{4}{c}{Interruption} \\
Method & BS-F1 $\uparrow$ &Task Acc $\uparrow$ & KIR $\uparrow$ & G-Eval $\uparrow$ & BS-F1 $\uparrow$ &Task Acc $\uparrow$ & KIR $\uparrow$ & G-Eval $\uparrow$ \\
\midrule
Naive LLM Agent & \textbf{0.866} & 0.72 & 0.38 & 2.96 & \textbf{0.855} & 0.03 & 0.53 & 2.92\\
FOCAL-GM & 0.855 & 0.68 & 0.42 & 3.41 & 0.852 & 0.80 & 0.30 & 2.45\\
FOCAL & 0.857 & 0.69 & \textbf{0.61} & \textbf{4.16} & 0.854 & \textbf{0.81} & \textbf{0.80} & \textbf{4.12}\\
\bottomrule
\end{tabular}%
%}
\end{table*}

\subsection{Case Study}
\label{sec:case_study}

To provide qualitative evidence for the quantitative improvements reported above, we examine one representative multi-task session and compare how the desktop stream is organized into task-level spans.

Figure~\ref{fig:case_study_stream} provides a visual overview of Session~258, which spans three coarse tasks: downloading an image from the web, editing it in Photoshop, and preparing slides in PowerPoint. FOCAL correctly recovers these three task-level spans from the raw activity stream, whereas the naive baseline fragments the same workflow into shorter, semantically shallower segments that fail to capture the compositional structure of the session.
This example is representative because it includes both cross-application transitions and a return to an earlier objective, two patterns that frequently occur in real desktop work.

The visual comparison shows a clear structural difference. Ground
truth and FOCAL both preserve three coherent task spans, whereas the
naive baseline breaks the same stream into many short segments. Once
this fragmentation occurs, semantically related actions are no longer
grouped under a stable task context, weakening task-level composition
and making later summaries less faithful.

This case study illustrates why task-isolated memory matters in
continuous desktop logging. By routing evidence into task-scoped
context before summary generation, FOCAL preserves higher-level
workflow structure across application switches instead of letting
brief interruptions overwrite the active narrative. This qualitative
pattern matches the gains in KIR and G-Eval reported in
Table~\ref{tab:quality_results}, showing that better task organization
leads directly to more faithful session summaries.
Notably, FOCAL does not simply equate every interface transition with
a task boundary: it preserves semantic continuity when the user
returns to an unfinished objective after a short detour, which is
precisely the failure mode that harms naive global summarization.

\section{Conclusion}

We present FOCAL, a privacy-first, on-device multi-agent logging
system for desktop activity summarization that instantiates filtered
on-device continuous activity logging through metadata-guided
filtering and task-isolated memory.
Experiments show that FOCAL makes desktop logging both efficient and
semantically reliable: on multi-task sessions it reduces total token
cost by 60.4\% versus a naive baseline while improving KIR from 0.38
to 0.61 and G-Eval from 2.96 to 4.16, and it remains robust under
frequent task switching.
FOCAL therefore provides a practical foundation for continuous,
resource-aware, and task-structured desktop logging.
More broadly, the results suggest that moving task-aware control ahead
of expensive visual reasoning is a viable design principle for
privacy-sensitive personal computing systems.
Current evidence is still limited to one hardware profile (Apple M4,
16\,GB) and one local model configuration (\texttt{qwen3:8b}).

Future work will extend evaluation to more model scales and hardware
settings, especially smaller local models with stronger structured
instruction following.
We also plan to study longer continuous sessions together with latency
and energy measurements to better characterize real on-device
deployment costs.
Another promising direction is to enrich the system from retrospective
logging to lightweight assistance, such as task resumption, progress
tracking, and privacy-preserving personal memory retrieval.

\bibliographystyle{ACM-Reference-Format}
\bibliography{sample-base}

\appendix

\end{document}